\documentclass[twocolumn,superscriptaddress,longbibliography,
	aps,prx,preprintnumbers,floatfix]{revtex4-2}

\usepackage[normalem]{ulem}
\usepackage{graphicx}
\usepackage{bm}
\usepackage{color}
\usepackage{epstopdf}
\usepackage{amsmath}
\usepackage{amssymb}
\usepackage{epstopdf}
\usepackage{lipsum}

\usepackage{float}

\usepackage{multirow}

\usepackage[urlcolor=blue,colorlinks=true,citecolor=blue,linkcolor=blue,pdfstartview={FitH},bookmarks=false]{hyperref}

\graphicspath{{fig/}{./fig/}{.}}

\begin{document}

\title{
Dynamical Study of the Origin of the Charge Density Wave \\ 
in $A$V$_{3}$Sb$_{5}$ ($A=$K, Rb, Cs) Compounds
}

\author{Andrzej~Ptok}
\email[e-mail: ]{aptok@mmj.pl}
\affiliation{\mbox{Institute of Nuclear Physics, Polish Academy of Sciences, W. E. Radzikowskiego 152, PL-31342 Krak\'{o}w, Poland}}

\author{Aksel~Kobia\l{}ka}
\affiliation{\mbox{Institute of Physics, Maria Curie-Sk\l{}odowska University, Plac Marii Sk\l{}odowskiej-Curie 1, PL-20031 Lublin, Poland}}

\author{Ma\l{}gorzata~Sternik}
\affiliation{\mbox{Institute of Nuclear Physics, Polish Academy of Sciences, W. E. Radzikowskiego 152, PL-31342 Krak\'{o}w, Poland}}

\author{Jan~\L{}a\.{z}ewski}
\affiliation{\mbox{Institute of Nuclear Physics, Polish Academy of Sciences, W. E. Radzikowskiego 152, PL-31342 Krak\'{o}w, Poland}}

\author{Pawe\l{}~T.~Jochym}
\affiliation{\mbox{Institute of Nuclear Physics, Polish Academy of Sciences, W. E. Radzikowskiego 152, PL-31342 Krak\'{o}w, Poland}}

\author{Andrzej~M.~Ole\'{s}}
\affiliation{\mbox{Institute of Theoretical Physics, Jagiellonian University,
Prof. Stanis\l{}awa \L{}ojasiewicza 11, PL-30348 Krak\'{o}w, Poland}}
\affiliation{Max Planck Institute for Solid State Research,
Heisenbergstrasse 1, D-70569 Stuttgart, Germany}

\author{Przemys\l{}aw~Piekarz$\,$}
\affiliation{\mbox{Institute of Nuclear Physics, Polish Academy of Sciences, W. E. Radzikowskiego 152, PL-31342 Krak\'{o}w, Poland}}

\date{\today}

\begin{abstract}
Systems containing the ideal kagome lattice can exhibit several distinct and novel exotic states of matter.
One example of such systems is a recently discovered $A$V$_{3}$Sb$_{5}$ ($A$ = K, Rb, and Cs) family of compounds.
Here, the coexistence of the charge density wave (CDW) and superconductivity is observed.
In this paper, we study the dynamic properties of the $A$V$_{3}$Sb$_{5}$ systems in context of origin of the CDW phase.
We show and discuss the structural phase transition from $P6/mmm$ to $C2/m$ symmetry that are induced by the presence of phonon soft modes.
We conclude that the CDW observed in this family of compounds is a consequence of the atom displacement, from the high symmetry position of the kagome net, in low-temperature phase.
Additionally, using the numerical {\it ab initio} methods, we discuss the charge distribution on the $A$V$_{3}$Sb$_{5}$ surface.
We show that the observed experimental 
stripe-like modulation of the surface, can be related to surface reconstruction and manifestation of the three dimensional $2 \times 2 \times 2$ bulk CDW.
Finally, the consequence of realization of the $C2/m$ structure on the electronic properties are discussed.
We show that the electronic band structure reconstruction and the accompanying modification of density of states correspond well to the experimental data.
\end{abstract}

\maketitle

\section{Introduction}
\label{sec:int}

Kagome systems are a very attractive venue to study exotic states in the condensed matter physics, due to the realization of several unusual features: topological electronic flat bands~\cite{ye.kang.18,lin.choi.18,sales.yan.19,meier.du.20,kang.fang.20}, anomalous Hall effect~\cite{chen.niu.14}, exotic type of superconductivity~\cite{ko.lee.09}, topological insulating phase~\cite{guo.franz.09}, frustrated magnetism~\cite{gardner.gingras.10,nussinov.vanderbrink.15}, chiral spin state~\cite{ohgushi.murakami.00}, quantum spin liquid state~\cite{han.hlton.12,zhou.kanoda.17}, or chiral phonons~\cite{chen.wu.19}.
An example of this type of systems is a recently discovered kagome-metals family $A$V$_{3}$Sb$_{5}$ \mbox{($A$ = K, Rb, and Cs)}~\cite{ortiz.gomes.19}, exhibiting the giant anomalous Hall effect~\cite{yang.wang.20,yu.wu.21,zheng.chen.21}.

The $A$V$_{3}$Sb$_{5}$ compounds were identified as nonmagnetic~\cite{kenney.ortiz.21} $\mathbb{Z}_{2}$ topological good 
metals~\cite{ortiz.sarte.21,ortiz.teicher.21b,setty.hu.21}, with weak electronic correlation effects~\cite{luo.gao.21,zhao.wu.21}.
As a consequence, the topological surface states can be observed, for instance in CsV$_{3}$Sb$_{5}$ just below the Fermi level~\cite{hu.teicher.21}.
The transport experiments confirm  the emergence of nontrivial band topology~\cite{fu.zhao.21}.
This opens a way for envisioning the new quantum devices, based on the topological properties of $A$V$_{3}$Sb$_{5}$ like, e.g., Josephson junctions with extremely long coupling (of at least $6$~$\mu$m in the case of KV$_{3}$Sb$_{5}$~\cite{wang.yang.20}).

One of the most interesting properties of the $A$V$_{3}$Sb$_{5}$ compounds is the coexistence of charge density wave (CDW) and superconductivity at low temperatures.
The transition to the CDW phase is observed at $T_\text{CDW}$ equal to $78$~K, $102$~K, and $94$~K for KV$_{3}$Sb$_{5}$~\cite{ortiz.sarte.21,li.zhang.21,zhu.yang.21}, RbV$_{3}$Sb$_{5}$~\cite{yin.tu.21,zhu.yang.21}, and CsV$_{3}$Sb$_{5}$~\cite{ortiz.gomes.19,yang.wang.20,ortiz.sarte.21,ortiz.teicher.21b,zhao.wang.21,yu.wu.21,liang.hou.21,mu.yin.21,yin.tu.21,song.zheng.21,li.zhang.21,wang.jiang.21,yu.wang.21,gupta.das.21,wang.kong.21}, respectively.
Further lowering of the temperature leads to the emergence of superconducting state at the critical temperature $T_{c}$, which is approximately equal to $0.93$~K, $0.92$~K, and $2.5$~K for KV$_{3}$Sb$_{5}$~\cite{ortiz.sarte.21}, RbV$_{3}$Sb$_{5}$~\cite{yin.tu.21}, and CsV$_{3}$Sb$_{5}$~\cite{ortiz.teicher.21b,mu.yin.21,ni.ma.21}, respectively.
Values of $T_{c}$ estimated from the electron-phonon coupling are lower than experimentally established ones~\cite{tan.liu.21}, which indicates an unconventional pairing mechanism~\cite{wu.schwemmer.21}.
Additionally, a general relatively weak dependence of the $T_{c}$ and $T_\text{CDW}$ on the thickness of the sample was reported for $A$V$_{3}$Sb$_{5}$ compounds~\cite{song.kong.21,song.ying.21,wang.yu.21}.
The imposed external hydrostatic pressure leads to the occurrence of double superconducting dome on the temperature--pressure phase diagram~\cite{zhu.yang.21}.
This behavior was observed in each of the $A$V$_{3}$Sb$_{5}$ family members, i.e., KV$_{3}$Sb$_{5}$~\cite{du.luo.21}, RbV$_{3}$Sb$_{5}$~\cite{wang.chen.21,du.luo.21b}, and CsV$_{3}$Sb$_{5}$~\cite{zhao.wang.21,yu.ma.21,chen.wang.21,chen.zhan.21,qian.christensen.21,zhang.chen.21,wang.kong.21}.
Generally, an increasing pressure clearly has a negative impact on the charge order causing a fast decrease of $T_\text{CDW}$, whereas the superconducting phase survives even at high pressures.

The three-dimensional (3D) CDW phase realized in $A$V$_{3}$Sb$_{5}$~\cite{ortiz.sarte.21,li.zhang.21,zhu.yang.21,yin.tu.21,zhu.yang.21,ortiz.gomes.19,yang.wang.20,ortiz.sarte.21,ortiz.teicher.21b,zhao.wang.21,yu.wu.21,liang.hou.21,mu.yin.21,yin.tu.21,song.zheng.21,li.zhang.21,wang.jiang.21,yu.wang.21,gupta.das.21}
is associated with the emergence of the $2 \times 2$ pattern of charge distribution on the surface, observed by the scanning tunnelling microscope (STM) experiments~\cite{jiang.yin.20,li.wan.21,zhao.li.21,chen.yang.21,liang.hou.21,li.zhao.21,wang.kong.21}.
However, it should be noted that the $4 \times 1$ stripe phase was also reported, e.g. in KV$_{3}$Sb$_{5}$~\cite{li.zhao.21}, RbV$_{3}$Sb$_{5}$~\cite{yu.xiao.21} or CsV$_{3}$Sb$_{5}$~\cite{li.wan.21,xu.yan.21,wang.kong.21}.
Nevertheless, recent studies suggest that the $4 \times 1$ structure can be associated with mesoscopic structural phase separation and does not exist in the bulk~\cite{li.jiang.21}.

Angle--resolved photoemission spectroscopy (ARPES) study does not give unequivocal information about the gap opening due to the CDW, e.g., the helium-lamp-based measurements report strong anisotropy of CDW gap opening around the K point~\cite{nakayama.li.21,luo.gao.21,wang.ma.21}, or a nearly isotropic gap behavior around the K point was found by the synchrotron-based study~\cite{lou.fedorov.21,kang.fang.21}.
Additionally, around the $\Gamma$ point, the gapless structure is reported~\cite{wang.ma.21}.
Nevertheless, the shape of the Fermi surface (FS) reported in many ARPES~\cite{kang.fang.21,luo.gao.21,hu.wu.21,li.zhang.21,hu.teicher.21,luo.peng.21,cai.wang.21,cho.ma.21} and {\it ab initio}~(DFT)~\cite{uykur.ortiz.21b,ortiz.teicher.21,luo.gao.21,fu.zhao.21,cho.ma.21,ortiz.teicher.21b,labollita.botana.21} studies support the idea that the emergence of the CDW is a consequence of the Peierls instability related to FS nesting near the saddle points at M~\cite{zhou.li.21,wang.liu.21}.

\paragraph*{Motivation.}---
The minimal model that describes the CDW realized in $A$V$_{3}$Sb$_{5}$ is focused only on the kagome lattice of V atoms~\cite{ortiz.teicher.21}.
Many papers assume that the CDW in a form of the {\it Star of David} (SD) or inverse SD (tri-hexagonal) structure~\cite{tan.liu.21,ortiz.teicher.21,christensen.birol.21} is a consequence of the kagome lattice distortion.
However, the observed CDW is not only associated with the kagome lattice~\cite{ortiz.teicher.21}, but also with the layer of Sb atoms~\cite{jiang.yin.20,liang.hou.21,shumiya.hossain.21}.
For example, the electronic band structure study of CsV$_{3}$Sb$_{5}$ shows that Sb atoms can play a role in realization of the CDW~\cite{tsirlin.fertey.21}.
Study of the
band structure of $A$V$_{3}$Sb$_{5}$ under pressure shows also the important role of the Sb bands~\cite{labollita.botana.21}.

The nuclear magnetic resonance (NMR) and nuclear quadrupole resonance (NQR) experiments also suggest an emergence of complex CDW phase.
In Ref.~\cite{luo.zhao.21}, authors discuss the NMR spectra of V in CsV$_{3}$Sb$_{5}$ and NQR spectra of Sb.
They observed that CDW is accompanied by an additional charge modulation in bulk below $\sim 40$~K.
Additionally, the NMR studies of Cs and V atoms in CsV$_{3}$Sb$_{5}$ report a displacement of Cs atoms for temperatures below $T_\text{CDW}$~\cite{song.zheng.21}.

Recent theoretical studies tried to explain the CDW formation by orbital current orders~\cite{denner.thomale.21,tan.liu.21,lin.nandkishore.21,wu.schwemmer.21,mielke.das.21,park.ye.21}, 
and realization of real and imaginary CDW scenario~\cite{lin.nandkishore.21}. 
For example, a competition between charge density order (CDO) and the charge bond order (CBO) was discussed with respect to the interaction within the kagome lattice~\cite{denner.thomale.21}.
However, these explanations completely ignore the interplay between the kagome net and the rest of the system, i.e., an additional Sb atom in the kagome plane.
As mentioned in the previous paragraphs, the CDW emerging in the $A$V$_{3}$Sb$_{5}$ compound family can have a much more complex nature than it was recently assumed.

In our paper, we discuss the emergence of the CDW phase from the dynamical point of view, i.e., in the context of the structural phase transition at $T_\text{CDW}$. 
Analysis of spectra measured using the inelastic x-ray scattering (IXS) and Raman spectroscopy of CsbV$_{3}$Sb$_{5}$ at different temperatures show that the P6/mmm probably  transforms to a lower symmetry structure with decreasing temperature ~\cite{ratcliff.hallett.21, wang.wu.21,wulferding.lee.21}. 
The existence of structural phase transition is also indicated by the theoretical studies of $A$V$_{3}$Sb$_{5}$~\cite{cho.ma.21,zhang.liu.21,tan.liu.21,subedi.21} which present the imaginary modes in the phonon dispersion relations.
What is more, even a simple SD or inverse SD deformation introduced ``by hand'' can lead to a dynamically stable structure, i.e., without imaginary phonon frequencies~\cite{tan.liu.21}. 
However, there is no clear evidence that these phases, i.e. SD or inverse SD, are a true ground states of the $A$V$_{3}$Sb$_{5}$ compounds.
Additionally, some theoretical studies show that the coupling between lattice distortions and CDW induces a weak, first order transition without a continuous phonon softening in $A$V$_{3}$Sb$_{5}$~\cite{miao.li.21}.

The group theory analysis shows that there are seventeen different structural distortions possible, due to the phonon instabilities at the M and L points in the parent $P6/mmm$ phase of $A$V$_{3}$Sb$_{5}$~\cite{subedi.21}.
The calculations of each distorted structures performed by Alaska Subedi~\cite{subedi.21} show that the $Fmmm$ gives the lowest energy for all $A$V$_{3}$Sb$_{5}$ family members.
However, the crystal structure optimized within the $Fmmm$ symmetry is unstable, showing the imaginary modes at the $\Gamma$ and Z points.

Our calculations of the dynamical properties of the $A$V$_3$Sb$_5$ compounds  performed  at two different temperatures $T=50$~K and $T=150$~K revealed a softenig of phonon modes at the M and L points at lower temperature for each compound.
Although, in KV$_3$Sb$_5$, all phonon dispersions are stable without imaginary frequencies, the weak decrease of both mode frequencies is still clearly visible.
RbV$_3$Sb$_5$ exhibits one imaginary mode at the L point but one frequency at M point is also reduced.
In CsV$_3$Sb$_5$ two soft modes at the L and M points were found.
Next, we show that crystal distortion defined by the combination of the polarization vectors from these two points leads to the stable $C2/m$ structure, which properties correspond well to the experimental results discussed above.

The paper is organized as follows.
The calculation methods are explained in Sec.~\ref{sec:met}. 
In Sec.~\ref{sec.num} we present the {\it ab initio} results -- we start with details of numerical calculations, then we discuss the dynamical properties of the basic $P6/mmm$ structure (Sec.~\ref{sec.dyn_prop}), its structural phase transtion to the $C2/m$ structure (Sec.~\ref{sec.transition}), the phonon density of states and Raman modes (Sec.~\ref{sec.dosy}). Next, we discuss the emergence of charge density wave using the STM simulations of the surface (Sec.~\ref{sec.cdw_surf}), and finally the electronic properties of both structures (Sec.~\ref{sec.el}).
We conclude our study in Sec.~\ref{sec.sum}.


\begin{figure*}[!t]
\centering
\includegraphics[width=\linewidth]{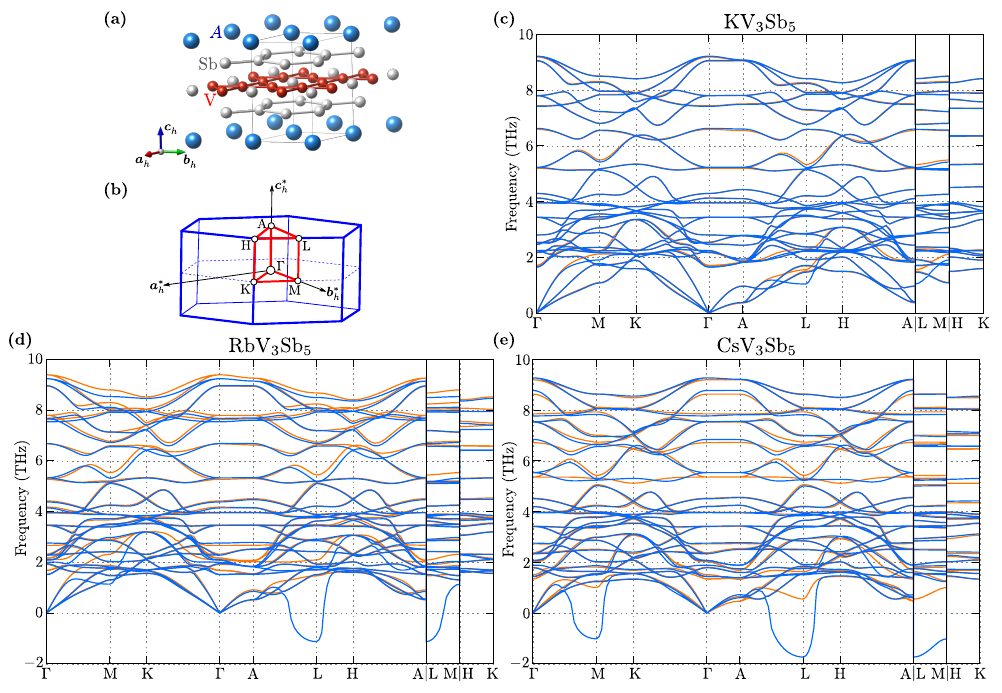}
\caption{
Common properties of the $A$V$_{3}$Sb$_{5}$ compounds with the $P6/mmm$ symmetry: (a) primitive unit cell (solid line) and 
(b)~the corresponding Brillouin zone. The remaining panels (c)-(e)
compare the phonon dispersions of these compounds obtained from the distribution of displacements for high and low temperature, i.e., $150$~K (orange line) and $50$~K (blue line), respectively.
In~the case of RbV$_{3}$Sb$_{5}$ and CsV$_{3}$Sb$_{5}$ the soft modes are observed.
\label{fig.band191}
}
\end{figure*}

\section{Calculation methods}
\label{sec:met}

The first-principles density functional theory (DFT) calculations were performed using the projector augmented-wave (PAW) potentials~\cite{blochl.94} implemented in 
the Vienna Ab initio Simulation Package ({\sc Vasp}) code~\cite{kresse.hafner.94,kresse.furthmuller.96,kresse.joubert.99}.
For the exchange-correlation energy the generalized gradient approximation (GGA) in the Perdew, Burke, and Ernzerhof (PBE) parametrization was used~\cite{pardew.burke.96}.
The energy cutoff for the plane-wave expansion was set to $350$~eV.
The van der Waals correction was included using the Grimme method~\cite{grimme.antony.10} implemented within {\sc Vasp}.
Optimizations of the structural parameters (lattice constants and atomic positions) in the primitive unit 
cell were performed using $16\times 16 \times 8$ ($8 \times 8\times 8$) {\bf k}--point grid in the case of hexagonal $P6/mmm$ (monoclinic $C2/m$) symmetry using the Monkhorst--Pack scheme~\cite{monkhorst.pack.76}.
As a convergence condition of the optimization loop, we took the energy change below $10^{-6}$~eV and $10^{-8}$~eV for ionic and electronic degrees of freedom. 
Symmetry of the structures were analyzed with {\sc FindSym}~\cite{stokes.hatch.05} and {\sc Seek-path}~\cite{hinuma.pizzi.17,togo.tanaka.18} packages.

The interatomic force constants (IFC) were obtained with the {\sc Alamode} software~\cite{tadano.gohda.14}, using the supercell technique [details about supercell construction can be found in the Supplemental Material (SM)~\footnote{See Supplemental Material at [URL will be inserted by publisher] for details about construction of the supercells and crystallographic data for obtained structures with $P6/mmm$ and $C2/m$ symmetries. Supplemental Material contain also additional numerical results and data}].
Calculations were performed for the thermal distributions of multi-displacement of atoms for given finite temperatures~\cite{hellman.abrikosov.11}, generated within {\sc hecss} procedure~\cite{jochym.lazewski.21}.
The energy and the Hellmann-Feynman forces acting on all atoms were calculated with {\sc Vasp} for twenty five different configurations of atomic displacements in the supercell.
In dynamical properties calculations, we included both harmonic and higher-order contributions to phonons.
The frequencies of the optical modes at the $\Gamma$ point and their activities were obtained within the {\it Parlinski--Li--Kawazoe} direct method~\cite{phonon1}
implemented in the {\sc Phonon} software~\cite{phonon2}.

\section{Numerical results and discussion}
\label{sec.num}

\subsection{Dynamical properties of $P6/mmm$}
\label{sec.dyn_prop}

\begin{figure}[!b]
\centering
\includegraphics[width=\linewidth]{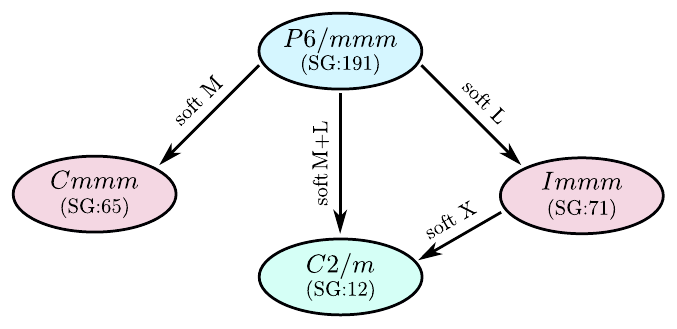}
\caption{
Diagram of symmetries generated by the soft modes (see labels). 
Initial symmetry $P6/mmm$ in low temperature can contain soft modes at M and L points, which leads to $Cmmm$ and $Immm$ symmetries, respectively.
However, a soft mode at the X point could lead to transition from $Immm$ to $C2/m$ symmetry. 
Similarly, the $C2/m$ symmetry could be induced jointly by both soft modes of $P6/mmm$ symmetry (from M and L points).
\label{fig.schemat}
}
\end{figure}

\begin{figure*}[!t]
\centering
\includegraphics[width=0.9\linewidth]{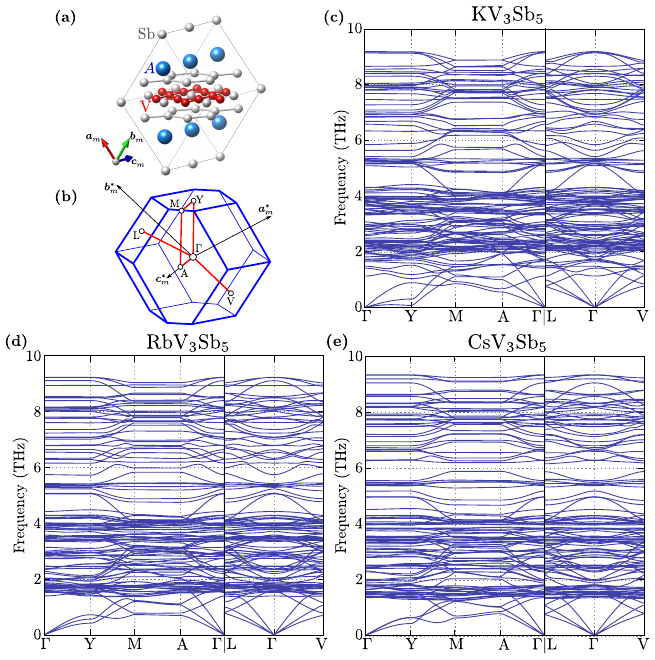}
\caption{Common properties of the $A$V$_{3}$Sb$_{5}$ compounds with the $C2/m$ symmetry: (a) primitive unit cell (solid line) and (b)~the corresponding Brillouin zone.
The remaining panels show phonon dispersions obtained for: (c) KV$_{3}$Sb$_{5}$, (d)~RbV$_{3}$Sb$_{5}$, and (e) CsV$_{3}$Sb$_{5}$. Note~that none of them exhibits any soft modes.
\label{fig.band12}
}
\end{figure*}

In the room temperature, the $A$V$_{3}$Sb$_{5}$ compounds crystallize in the the hexagonal $P6/mmm$ symmetry. 
Crystal structure, shown in Fig.~\ref{fig.band191}(a), contains the ideal kagome net of $A$ atoms (decorated by one Sb atom), sandwiched by two honeycomb nets of Sb atoms (details of the crystal structure are given in the SM~\cite{Note1}).
Taking the generic common description presented in the SM~\cite{Note1}, system possesses only three structural degrees of freedom ($a$ and $c$ lattice constants, and $z_\text{Sb}$-coordinate describing the position of Sb atom in honeycomb layer).
The irreducible representations (IR) at the $\Gamma$ point are: $A_\text{2u} + E_\text{1u}$ for acoustic modes, and $A_\text{1g} + 3A_\text{2u} + B_\text{1g} + B_\text{1u} + 2 B_\text{2u} + 2 E_\text{2u} + E_{2g} + 4 E_\text{1u} + E_\text{1g}$ for optical modes.
Only $A_\text{1g}$, $E_\text{2g}$, and $E_\text{1g}$ modes are Raman active, while $A_\text{2g}$ and $E_\text{1u}$ modes are infra-red active.

To examine the relationship between the CDW phase and the lattice dynamics of 
the studied compounds,  we calculated the phonon dispersion relations at two temperatures strictly related with the emergence of the CDW. 
The dispersion curves obtained for $A$V$_{3}$Sb$_{5}$  systems at 150 K (above $T_\text{CDW}$) and 50 K (below $T_\text{CDW}$) are demonstrated in Fig.~\ref{fig.band191} as orange and blue lines, respectively.
In the case of heavy $A$ ions, i.e. in RbV$_{3}$Sb$_{5}$ and CsV$_{3}$Sb$_{5}$, we observed the imaginary modes for $T < T_\text{CDW}$.
In the first case, the imaginary mode at the L point and the small frequency softening at the M point are observed, while in the second, two imaginary modes are found at the M and L points.
Imaginary frequency modes lead to the structural phase transition, which will be discussed in more detail in the next subsection.
We did not observe any imaginary modes in KV$_{3}$Sb$_{5}$ compound, however, the frequencies of both modes, at the M and L points, are slightly reduced at 50 K.
The imaginary frequencies are present when the phonon dispersion relations are calculated at T = 0~K ~\cite{zhang.liu.21,tan.liu.21,cho.ma.21} thus we assume that all three compounds transform to the lower symmetry phase in the same way.

\subsection{Structural phase transition to $C2/m$}
\label{sec.transition}

Symmetry analyses of soft modes gives information about the possible stable structures (Fig.~\ref{fig.schemat}) for studied systems.
First of all, the soft mode at the M point can induce a transition to the $Cmmm$ symmetry (space group $65$), while the soft mode at the L point 
to the $Immm$ symmetry (space group $71$). 
It is important to note, that the lowest
soft mode typically leads to the most energetically favorable structure. 
In our case, this corresponds to the soft modes at the L point, which not only change the atom positions, but also generate distortion of the lattice.
What is more, dynamical study of the obtained structure with the $Immm$ symmetry leads to another soft mode (at the X point) resulting in a stable $C2/m$ structure.
This structure can be also found as a result of simultaneous condensation of both soft modes from the M and L points.

Structure with the $C2/m$ symmetry [presented in Fig.~\ref{fig.band12}(a)] does not exhibit any soft modes for any of the $A$V$_{3}$Sb$_{5}$ compounds~\footnote{Notice, that the distortion introduced by-hand in form of \textit{Star of David} or inverse \textit{Star of David} can give positive phonon spectra, see Ref.~\cite{wang.kong.21} and~\cite{tan.liu.21}.}.
Phonon dispersions [Figs.~\ref{fig.band12}(c)-(e)] are characterized by many nearly flat-bands, as a consequence of the Brillouin zone folding (in relation to the basic $P6/mmm$ symmetry).
The IRs at the $\Gamma$ point are: $A_\text{u} + 2 B_\text{u}$ for acoustic modes, and $26 A_\text{g} + 23 A_\text{u} + 22 B_\text{g} + 34 B_\text{u}$ for optical modes.
$A_\text{g}$, and $B_\text{g}$ modes are Raman active, while $A_\text{u}$ and $B_\text{u}$ modes are infra-red active.

Now, we will discuss the atomic displacements during the transition from the $P6/mmm$ to $C2/m$ phase because the analysis of the atoms rearrangement  may shed new light on the CDW phase of $A$V$_{3}$Sb$_{5}$. 
Both phases have a layered framework and the corresponding layers can be easily 
compared [cf.~Fig.~\ref{fig.band191}(a) and Fig.~\ref{fig.band12}(a)].
We will discuss this question on the example of CsV$_{3}$Sb$_{3}$.
First, the $A = \text{Cs}$ atoms change position in the direction perpendicular to the layers (along initial $c_{h}$ direction).
The V atoms, on the contrary, are moved only in the plane of the layer.
Finally, Sb atoms have rather complex behavior--- some of them remain at original positions, while the rest of them is shifted (mostly in $c_{h}$ direction).
The biggest displacements are observed in the case of V atoms (even up to $0.10$~\AA) and some Sb atoms (smaller than $0.05$~\AA).

\begin{figure}[!b]
\centering
\includegraphics[width=.98\linewidth]{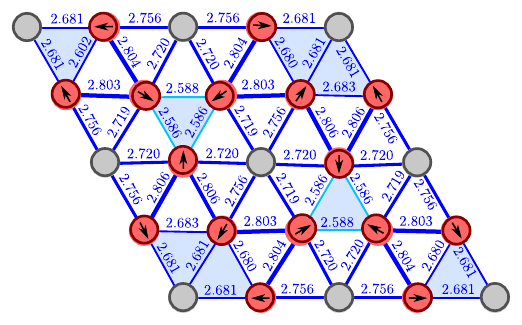}
\caption{
Calculated distances between atoms in the kagome net layer for CsV$_{3}$Sb$_{5}$.
V and Sb atoms are represented by red and gray circles, respectively.
Position of atoms for $P6/mmm$ structure, is given by colored circles, while for $C2/m$ by solid line circles.
The blue lines width corresponds to the distance between atoms, which is also directly referenced by its value (in~\AA).
``Ideal'' distance between atoms in $P6/mmm$ is $2.718$~\AA.
Black arrows show the direction of displacements.
\label{fig.vpalne}
}
\end{figure}

As far as the features of CDW phase are concerned, the most important one is the layer containing V atoms forming kagome net in the $P6/mmm$ structure (or distorted kagome-like net in $C2/m$).
In Fig.~\ref{fig.vpalne} we present the differences between atomic positions in the ideal and distorted layer.
There, the colored circles correspond to atomic sites in the ideal kagome net layer (the distance between atoms is $2.718$~\AA), while solid line circles denote positions of atoms in the distorted lattice of $C2/m$ phase.
The Sb atoms in the distorted kagome-like layer actually conserve their positions. 
In contrast to that, the V atoms change their positions significantly.
We can define three centers of atomic ``attraction'' (marked by shaded blue triangles in Fig.~\ref{fig.vpalne}). 
Two of them, form isoscales triangles of V atoms, with the distance between atoms equal to $\sim 2.58$~\AA.
Third center is realized in the form of the hexagon of V atoms and centered on Sb atom.
In this case, the distance between atoms is equal to $\sim 2.68$~\AA.
As for the rest, the distances between atoms are unchanged or bigger than in the initial $P6/mmm$ structure (detailed values are shown in Fig.~\ref{fig.vpalne}).
Summarizing the above, the low-temperature $C2/m$ structure bears the similarity to the inverse SD (tri-hexagonal) structure.


\subsection{The phonon density of states and Raman scattering}
\label{sec.dosy}

\begin{figure}[!t]
\centering
\includegraphics[width=0.91\linewidth]{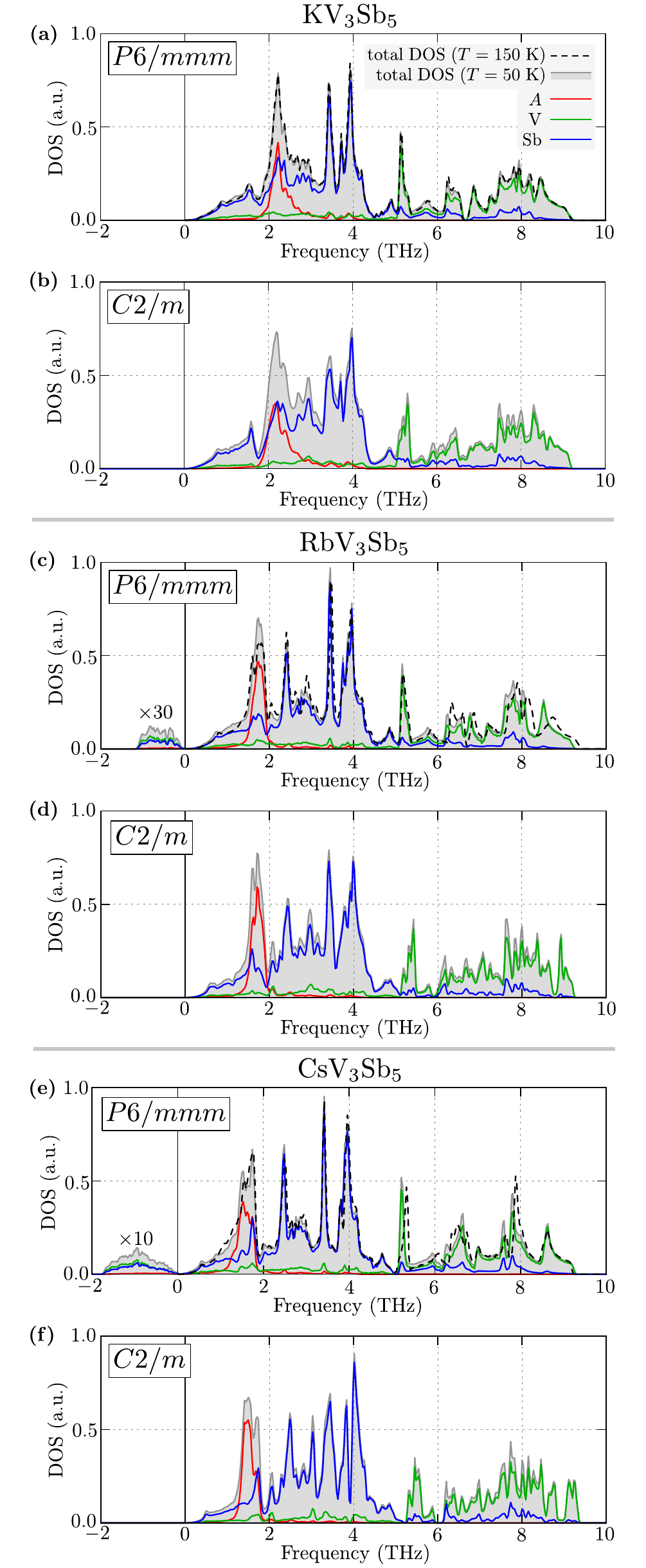}
\caption{
Phonon density of states (DOS) for $A$V$_{3}$Sb$_{5}$ family members with different symmetries (as labeled).
The grey background denotes the total density of states (for $T = 50$~K), whereas  the red, green, and blue lines to the partial DOS of $A$, V, and Sb atoms, respectively.
Additionally, for $P6/mmm$ symmetry, as the black dashed line we show total DOS at $T = 150$~K.
Imaginary frequencies, shown as negative values on (c) and (f), were magnified as a guide-for-eye 30 and 10 times, respectively.
\label{fig.dos}
}
\end{figure}

Now, we shortly discuss the phonon density of states (DOS) shown in Fig.~\ref{fig.dos}.
For each of the discussed structures (independently of the symmetry), the highest frequency modes (above $5$~THz) correspond mostly to vibrations of V atoms (solid green line).
Near-flat-bands, located mostly in the middle range of frequencies (from $2$~THz to $5$~THz) are mainly associated with Sb atoms (solid blue lines).
Finally, vibrations of alkali metals are located around $\sim 1.75$~THz, and form a clear peak in the DOS (solid red line).
In the case of the soft mode DOS, spectral weights clearly show  contributions from Sb and V atoms [see Figs.~\ref{fig.dos}(c) and \ref{fig.dos}(e)].
Additionally, for the $C2/m$ structures, we observed a lifting of degeneracy of the band in relation to the $P6/mmm$ symmetry, in the form of multi-peak structure. 
In practice, this DOS modification is observed in the whole range of frequencies (especially for the contributions that originate from V and Sb atoms).

The frequencies of the Raman modes with their irreducible representations calculated for CsV$_3$Sb$_5$ with the $P6/mmm$ and $C2/m$ symmetries are presented in Tab.~\ref{tab.ir} in SM~\cite{Note1}. 
The Raman active modes in the $P6/mmm$ structure are present only within the honeycomb Sb sublattices -- the $A_{1g}$ mode represents the out-of-plane vibrations, 
while $E_{1g}$ and $E_{2g}$ are the in-plane modes.
The $A_{1g}$ mode is found at $4.2$~THz, in good agreement with the experiments~\cite{li.zhang.21,ratcliff.hallett.21,wang.wu.21}.
The frequency of the $E_{2g}$ mode ($3.9$~THz) is overestimated relative to experimental values, $3.63$~THz~\cite{li.zhang.21} and $3.55$~THz~\cite{wulferding.lee.21}, 
while the lowest $E_{1g}$ mode has lower frequency $2.27$~THz than the measured value $2.66$~THz~\cite{li.zhang.21}.
Our calculations cannot explain the peak at $\sim 4.6$~THz observed close to $T_\text{CDW}$~\cite{li.zhang.21}, which may be connected with the
CDW fluctuations induced by the laser pumping.
After phase transition to the $C2/m$ structure, the Raman active modes are also associated  with other atoms (i.e., Cs and V).
In the low-symmetry phase, besides the $A_g$ mode at $4.1$~THz, two additional modes around $1.3$~THz and $3.1$~THZ were observed in the pump-probe experiments~\cite{ratcliff.hallett.21,wang.wu.21}. 
They can be identified as the symmetric $A_\text{g}$ modes found here at $1.4$~THz and $3.1$~THz (see Table~\ref{tab.ir} in SM~\cite{Note1}).
The lower mode appears just below $T_\text{CDW}$, while the higher mode is observed below $T^{\ast}\sim 60$~K. 
It may indicate the existence of the intermediate phase between $T^{\ast} < T < T_\text{CDW}$ with a different symmetry -- one of the allowed by the soft modes symmetry~\cite{subedi.21}.

\begin{figure}[!b]
\centering
\includegraphics[width=\linewidth]{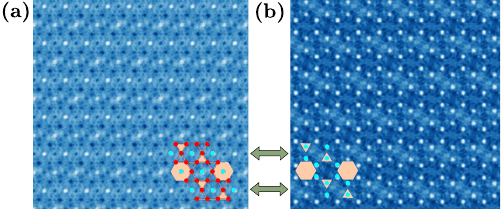}
\caption{
The DFT-generated STM image for CsV$_{3}$Sb$_{5}$ surface: (a) terminated by the distorted kagome-like plane and (b)~terminated by Sb plane.
Positions of the V and Sb atoms are marked by red and cyan dots, respectively.
Marked orange triangles and hexagon correspond to the shortest distances between V atoms in kagome-like plane (cf. Fig.~\ref{fig.vpalne}). 
Arrows indicate horizontal lines with identical stripe-like charge distribution. 
\label{fig.stm}
}
\end{figure}

\subsection{Surface charge density wave}
\label{sec.cdw_surf}

The CDW phase in $A$V$_{3}$Sb$_{5}$ was visualized employing the STM experiment many times~\cite{liang.hou.21,chen.yang.21,li.wan.21,zhao.li.21,chen.yang.21,wang.jiang.21,liang.hou.21,xu.yan.21,jiang.yin.20,li.zhao.21,shumiya.hossain.21}.
In the case of KV$_{3}$Sb$_{5}$, a topography of the Sb layer shows the realization of the surface $2\times2$ modulation~\cite{jiang.yin.20,li.zhao.21}.
Similar results were obtained for the Sb-terminated surface of RbV$_{3}$Sb$_{5}$~\cite{shumiya.hossain.21}.
However, in this case the additional stripe-like 
surface structure can be observed in the STM image.
Finally, the $2\times2$ CDW and $4\times1$ stripe superstructures were reported also in the case of the CsV$_{3}$Sb$_{5}$  surface~\cite{liang.hou.21,chen.yang.21,wang.jiang.21,li.wan.21,zhao.li.21,xu.yan.21}.

There may be a few reasons for the formation of observed CDW structures.
Firstly, the atomic displacements during transition from $P6/mmm$ to $C2/m$ symmetry can generate the CDW phase as it is shown earlier in Sec.~\ref{sec.transition}.
Secondly, the surface reconstruction can play an important role 
in the emergence of the CDW pattern in the STM experiments.
To explore this issue, we performed the DFT calculations to simulate STM images 
for a low-temperature monoclinic phase of CsV$_{3}$Sb$_{5}$ 
(Fig.~\ref{fig.stm}).
In the figure, among the complex charge modulations we can identify a clear unidirectional stripes (pointed out by arrows) running along one of the lattice directions (horizontal lines). 
For clarity, we mark also positions of V and Sb atoms by red and cyan dots, respectively, and positions of the adjacent V atoms in kagome-like plane by orange triangles and hexagons (cf.~Fig.~\ref{fig.vpalne}).

In the case of V kagome-like plane termination [shown in Fig.~\ref{fig.stm}(a)], we observe a stripe-like structure corresponding to the pattern realized by the adjacent V atoms (mostly marked hexagonal shapes).
This finding agrees with the previous observation that the CDW phase in this plane is realized by the bonding of vanadium $d_{xz}/d_{yz}$ orbitals~\cite{wang.liu.21}.
Certain refinement can be given also by the shifts of the Sb atoms out of this plane since the realization of the vacuum state from one-site of this plane leads to the small displacement of the Sb atoms in the out-of-plane (vacuum) direction.
And then the positions of V atoms are modified, reconstructing a nearly-ideal kagome lattice.

In the case of the Sb-terminated surface [see Fig.~\ref{fig.stm}(b)], we also observe  the stripe-like structure.
Emergence of the CDW on this surface indirectly corresponds to the inverted SD structure.
Here, the surface charge distribution is governed mainly by the $p$ orbitals of Sb atoms.
However, as we show earlier, the distorted kagome-like plane disturbs the positions of the Sb atoms of neighboring layers, so consequently, the influence of 3D (bulk) CDW is also noticeable in this surface.
Similar effect was observed by studying the surface states of Bi$_{2}$Se$_{3}$ where the surface charges were modified by the vacancies or atom substitution in the interior of bulk material~\cite{kim.ye.11,wang.xu.11,jurszyszyn.sikora.20}, due to realization of long-range orbital hybridization~\cite{ptok.kapcia.20b}.
Concluding, the observed stripe-like superstructure is related to a subtle effects of the surface reconstruction and the atomic orbitals hybridization rather than an exact 3D CDW emerging in the $A$V$_{3}$Sb$_{5}$.


\subsection{Electronic properties}
\label{sec.el}

\begin{figure}[!t]
\centering
\includegraphics[width=\linewidth]{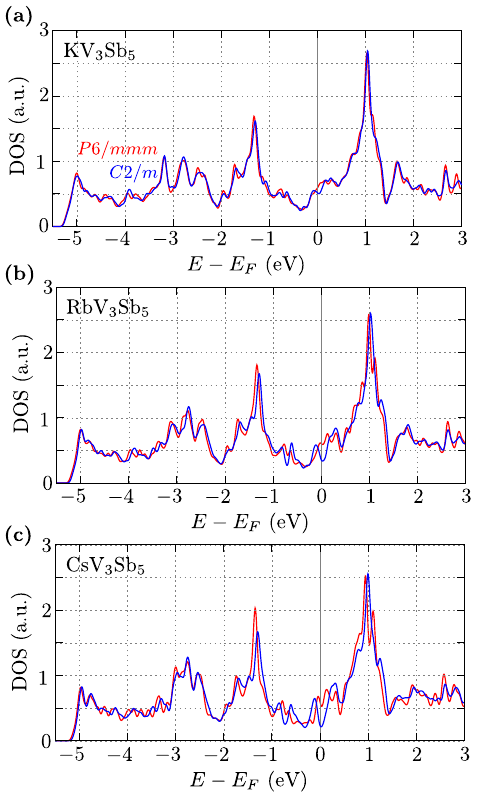}
\caption{
Comparison between the electronic density of states calculated for $P6/mmm$ (red line) and $C2/m$ (blue line) phases of $A$V$_{3}$Sb$_{5}$. 
Rows from top to bottom correspond to $A$ = K, Rb, and Cs, respectively.
\label{fig.el_dos}
}
\end{figure}

The theoretical study of the optical conductivity in CsV$_{3}$Sb$_{5}$~\cite{uykur.ortiz.21,uykur.ortiz.21b} shows that the phase transition to the CDW phase is manifested as the transfer of spectral weight towards the higher energies, indicating a partial gap opening and moreover, the reduction in the density of states at the Fermi level.
Indeed, the spectral weight suppression around the Fermi level was also observed  experimentally, in particular in KV$_{3}$Sb$_{5}$~\cite{jiang.yin.20}, RbV$_{3}$Sb$_{5}$~\cite{yu.xiao.21} or CsV$_{3}$Sb$_{5}$~\cite{liang.hou.21}.
In this context, van Hove singularities (VHS) were also studied in many papers~\cite{cho.ma.21,hu.wu.21}.
For example, in the case of CsV$_{3}$Sb$_{5}$ multiple VHS, emerging from $d_{xz}$/$d_{yz}$ and $d_{xy}$/$d_{x2-y2}$ kagome bands~\cite{kang.fang.21}, 
coexist near the Fermi level.
Additionally, the doping and pressure can tune the VHS closer to the Fermi level, which can also result in different Fermi surface instabilities~\cite{labollita.botana.21}.
Also, the band reconstruction due to system folding and a surface induced orbital-selective shift of the electron energy band were observed in ARPES experiments, e.g. in KV$_{3}$Sb$_{5}$~\cite{luo.gao.21} or in CsV$_{3}$Sb$_{5}$~\cite{luo.peng.21,ortiz.teicher.21}.

Now, we shortly discuss the electronic properties of the $A$V$_{3}$Sb$_{5}$ in the context of its band structure (Fig.~\ref{fig.el_band} in the SM~\cite{Note1}) and DOS (Fig.~\ref{fig.el_dos}).
In the case of the $P6/mmm$ symmetry (top panels in Fig.~\ref{fig.el_band} in the SM~\cite{Note1}), 
electronic band structures obtained for the primitive cell are in good agreement with the previous studies.
In each case, the band structure is formed mostly by V $3d$ orbitals~\cite{wang.liu.21,labollita.botana.21,zhao.wu.21,tsirlin.fertey.21}.
Realization of the ideal kagome lattice by V atoms leads to the occurrence of a nearly-flat band around $1$~eV, and the corresponding peak in electronic DOS (Fig.~\ref{fig.el_dos}).
At the K and M points, the Dirac points and saddle points are observed around the Fermi level, respectively.
This result is consistent with the previous {\it ab initio} studies of these compounds~\cite{ortiz.gomes.19,song.ying.21,chen.zhan.21,labollita.botana.21,zhou.li.21,wang.liu.21,hu.wu.21,kang.fang.21}.

Emergence of the 3D CDW within the $2\times 2\times 2$ supercell of $P6/mmm$ system leads to the above mentioned folding of the Brillouin zone and thus induces the band structure reconstruction (middle panels on Fig.~\ref{fig.el_band} in the SM~\cite{Note1}).
In the $C2/m$ band structure, new hole-like structures can be identified (marked by pink dotted lines).
Indeed, this type of band structure reconstruction is observed in the ARPES experiments~\cite{luo.peng.21,luo.gao.21}.
Furthermore, emergence of the similar structures can be observed for the $C2/m$ supercell (bottom panels in Fig.~\ref{fig.el_band} in the SM~\cite{Note1}).
In this case, additional effects of the band degeneracy lifting in high symmetry points can be observed. 
As we can see, in the case of both $P6/mmm$ 
and $C2/m$ 
supercells, their band structure looks similar. 
Remarkably, such an effect was reported previously~\cite{tang.ono.21,subedi.21} in the case of other supercells (e.g. with SD or inverted SD pattern introduced by-hand to the system).

Lowering the symmetry and lifting the band degeneracy (also at the saddle point) around the Fermi level generates a 
V-shape gap in the DOS spectrum exactly at the Fermi level (see Fig.~\ref{fig.el_dos}).
Similar effect was reported previously -- it influences the CDW phase.
Here, we clearly show that this is a consequence of the structural phase transition.
However, emergence of the additional CDW order should also lead to gap opening.
Therefore, this particular aspect of this study is still under debate.


\section{Summary and conclusion}
\label{sec.sum}

In this paper we have discussed the origin of the charge density wave in the $A$V$_{3}$Sb$_{5}$ compounds from the microscopic point of view using the {\it ab initio} study of their dynamical properties.
First, we showed that the structural phase transition at $T_\text{CDW}$ can be realized due to symmetry breaking from the $P6/mmm$ to $C2/m$.
Next, we demonstrated that the atomic displacement during this transition leads to emergence of the inverse \textit{Star of David} pattern.
This mechanism can be the source of the 3D $2\times2\times2$ charge density wave observed in these compounds.
Additionally, the $4\times1$ stripe superstructure observed experimentally on the surface of $A$V$_{3}$Sb$_{5}$ family can be a consequence of the surface reconstruction.

In our study, we used a thermal multi-displacement technique to find high-order phonon contributions.
Notably, lowering symmetry and stabilization of the system with the $C2/m$ symmetry is a consequence of the soft modes in the fundamental crystal structure with $P6/mmm$ symmetry.
The first-order structural transition can occur directly due to the synergy between soft modes from the M and L points.
We have found that, in practice, all of the atoms are shifted from their original high symmetry positions, which suggests that the emergence of more 
complex charge density wave phase is more likely than it was previously assumed.
Stemming from this, analyses of the atoms displacement confirm the appearance of the inverse \textit{Star of David} pattern, within the distorted kagome-like layer.

Possible manifestation of the $C2/m$ phase in $A$V$_{3}$Sb$_{5}$ provides an opportunity to observe experimentally the emergence of charge density wave phase 
by Raman active modes in the phonon spectrum. Indeed, it was recently observed for $T < T_\text{CDW}$.
Experimental observations of the folding of the Brillouin zone and the reconstruction of electronic band structure were also reproduced. 
In addition, lifting of the electronic band degeneracy leads to modification of the DOS at the Fermi level, in agreement with experimental results.


\begin{acknowledgments}
Some figures in this work was rendered using {\sc Vesta}~\cite{momma.izumi.11}.
This work was supported by National Science Centre (NCN, Poland) under Projects No.
2017/24/C/ST3/00276 (A.P.), 
2018/31/N/ST3/01746 (A.K.),
2016/23/B/ST3/00839 (A.M.O.),
and
2017/25/B/ST3/02586 (P.P.).
In addition, A.P. appreciates funding in the frame of scholarships of the Minister of Science and Higher Education of Poland for outstanding young scientists (2019 edition, No.~818/STYP/14/2019).
\end{acknowledgments}

\bibliography{biblio}

\newpage

\clearpage
\newpage

\onecolumngrid

\begin{center}
  \textbf{\Large Supplemental Material}\\[.2cm]
  \textbf{\large Dynamical Study of the Origin of the Charge Density Wave \\ [.2cm]
in $A$V$_{3}$Sb$_{5}$ ($A=$K, Rb, Cs) Compounds}\\[.2cm]
  Andrzej Ptok,$^{1}$ Aksel Kobia\l{}ka,$^{2}$ Ma\l{}gorzata~Sternik,$^{1}$ 
  Pawe\l{} T. Jochym,$^{1}$ \\ [.1cm] Jan \L{}a\.{z}ewski,$^{1}$ Andrzej M. Ole\'{s},$^{3,4}$ and Przemys\l{}aw Piekarz$\,^{1}$\\[.2cm]
  {\itshape
  	\mbox{$^{1}$Institute of Nuclear Physics, Polish Academy of Sciences, W. E. Radzikowskiego 152, PL-31342 Krak\'{o}w, Poland}\\
  	\mbox{$^{2}$Institute of Physics, Maria Curie-Sk\l{}odowska University, Plac Marii Sk\l{}odowskiej-Curie 1, PL-20031 Lublin, Poland}\\
  	\mbox{$^{3}$Institute of Theoretical Physics, Jagiellonian University,
Prof. Stanis\l{}awa \L{}ojasiewicza 11, PL-30348 Krak\'{o}w, Poland}\\
    \mbox{$^{4}$Max Planck Institute for Solid State Research,
Heisenbergstrasse 1, D-70569 Stuttgart, Germany}
  }
\\
(Dated: \today)
\\[1cm]
\end{center}

\setcounter{equation}{0}
\renewcommand{\theequation}{S\arabic{equation}}
\setcounter{figure}{0}
\renewcommand{\thefigure}{S\arabic{figure}}
\setcounter{section}{0}
\renewcommand{\thesection}{S\arabic{section}}
\setcounter{table}{0}
\renewcommand{\thetable}{S\arabic{table}}
\setcounter{page}{1}

\onecolumngrid

In this Supplemental Material we present additional results, in particular concerning:
\begin{itemize}
\item crystallographic data obtained for structures in $P6/mmm$ (Table~\ref{tab.latt191}) and $C2/m$ symmetries (Table~\ref{tab.latt12}),
\item information about the construction of the conventional cell and supercell used in calculations (Section~\ref{sec.cells}),
\item comparison of the electronic band structure for  $A$V$_{3}$Sb$_{5}$ supercells with different symmetries 
(Fig.~\ref{fig.el_band}),
\item the frequencies of the optical modes at the $\Gamma$ point with the indicated
Raman (R) and infrared (Ir) activities for CsV$_{3}$As$_{5}$ in $P6/mmm$ and $C2/m$ phases (Table~\ref{tab.ir}).
\end{itemize}

\section{Construction of conventional cell and supercell}
\label{sec.cells}

Let us start with the introduction of the basis $({\bm a}_{t},{\bm b}_{t},{\bm c}_{t})$ and $({\bm a}_{m},{\bm b}_{m},{\bm c}_{m})$, denoting the lattice vectors of the primitive unit cells for the tetragonal $P6/mmm$ and monoclinic $C2/m$ structure.
In the case of the $P6/mmm$ primitive unit cell it is equivalent to the conventional cell. 
Contrary to this, the conventional cell for $C2/m$ can be obtained by the following transformation:
\begin{equation}
\left( \begin{array}{rcc}
1 & 1 & 0 \\ 
-1 & 1 & 0 \\ 
0 & 0 & 1
\end{array}  \right) 
\left( \begin{array}{c}
{\bm a}_{m} \\ 
{\bm b}_{m} \\ 
{\bm c}_{m}
\end{array} \right) = \left( \begin{array}{c}
{\bm a}_{m}^{c} \\ 
{\bm b}_{m}^{c} \\ 
{\bm c}_{m}^{c}
\end{array} \right) .
\end{equation}
Values of the lattice vectors in this convention are presented directly in Table~\ref{tab.latt12}.

In the phonon calculation we used the supercell  with relatively large lattice vectors to minimize the self-interaction between atom and their simulated picture obtained from periodic boundary condition within DFT.
In the case of the $P6/mmm$ symmetry, we took the $3 \times 3 \times 2$ supercell, while for the $C2/m$ symmetry a supercell was given by a transformation:
\begin{equation}
\left( \begin{array}{ccc}
1 & 0 & 1 \\ 
0 & 1 & 0 \\ 
1 & 0 & 2
\end{array}  \right) 
\left( \begin{array}{c}
{\bm a}_{m}^{c} \\ 
{\bm b}_{m}^{c} \\ 
{\bm c}_{m}^{c}
\end{array} \right) = \left( \begin{array}{c}
{\bm a}_{m}^{s} \\ 
{\bm b}_{m}^{s} \\ 
{\bm c}_{m}^{s}
\end{array} \right) ,
\end{equation}
what gives the lattice vectors with length in range $\sim 18.5$~\AA.

\newpage

\begin{table*}[!pt]
\caption{
Lattice constants and atoms positions of $A$V$_{3}$Sb$_{5}$ for the $P6/mmm$ symmetry. Experimental data from Ref.~\cite{ortiz.gomes.19} were obtained for $T = 300$~K.
}
\begin{ruledtabular}
\begin{tabular}{ccccccc}
symmetry & \multicolumn{6}{c}{$P6/mmm$} \\
 \hline
based atom $A$ & \multicolumn{2}{c}{K} & \multicolumn{2}{c}{Rb} & \multicolumn{2}{c}{Cs} \\
& experimental~\cite{ortiz.gomes.19} & theoretical & experimental~\cite{ortiz.gomes.19} & theoretical & experimental~\cite{ortiz.gomes.19} & theoretical \\
 \hline
a (\AA) & 5.4818 & 5.4028 & 5.4715 & 5.4154 & 5.4949 & 5.4371 \\
b (\AA) & 5.4818 & 5.4028 & 5.4715 & 5.4154 & 5.4949 & 5.4371 \\
c (\AA) & 8.9477 & 8.6885 & 9.0733 & 8.8770 & 9.3085 & 9.0520 \\
V (\AA$^{3}$) & 232.86 & 219.64 & 235.24 & 225.45 & 243.41 & 231.74  \\
 \hline
$A$ (1a) & $\rightarrow$ & $\rightarrow$ & $\rightarrow$ & $\rightarrow$ & $\rightarrow$ & (0,0,0) \\
V (3g) & $\rightarrow$ & $\rightarrow$ & $\rightarrow$ & $\rightarrow$ & $\rightarrow$ & (1/2,0,1/2) \\
Sb (1b) & $\rightarrow$ & $\rightarrow$ & $\rightarrow$ & $\rightarrow$ & $\rightarrow$ & (0,0,1/2) \\
Sb (4h) & $\rightarrow$ & $\rightarrow$ & $\rightarrow$ & $\rightarrow$ & $\rightarrow$ & (1/3,2/3,$z_\text{Sb}$) \\ 
$z_\text{Sb}$ & 0.7532 & 0.76158 & 0.7498 & 0.75509 & 0.7421 & 0.74919 \\
\end{tabular}
\end{ruledtabular}
\label{tab.latt191}
\end{table*}

\begin{table*}[!pt]
\caption{
Lattice constants and atoms positions of $A$V$_{3}$Sb$_{5}$ for the $C2/m$ symmetry. 
}
\begin{ruledtabular}
\begin{tabular}{cccc}
symmetry & \multicolumn{3}{c}{$C2/m$} \\
 \hline
based atom $A$ & K & Rb & Cs \\
 \hline
a (\AA) & 10.8047 & 10.8299 & 10.8538 \\
b (\AA) & 17.3850 & 17.7788 & 18.3590 \\
c (\AA) & 10.8096 & 10.8299 & 10.8643 \\
V (\AA$^{3}$) & 1758.72 & 1807.89 & 1875.45  \\
$\alpha$ ($\deg$) & $\rightarrow$ & 90 & $\leftarrow$ \\
$\beta$ ($\deg$) & $\rightarrow$ & 119.975 & $\leftarrow$ \\
$\gamma$ ($\deg$) & $\rightarrow$ & 90 & $\leftarrow$ \\
 \hline  \hline
$A$ (4g) & $\rightarrow$ & (0,1/4,0) & $\leftarrow$ \\
$A$ (4h) & (0,0.2493,1/2) & (0,0.2490,1/2) & (0,0.2489,1/2) \\
 \hline
V (4i) & (0.00112,00.2519) & (0.0006,0,0.2529) & (0.0001,0,0.2535) \\
V (4i) & (0.2485,0,1/2)    & (0.2473,0,1/2)    & (0.2466,0,1/2) \\
V (4i) & (0.2493,0,0.7481) & (0.2477,0,0.7471) & (0.2466,0,0.7465) \\
V (4i) & (0.2514,0,0.2479) & (0.2531,0,0.2466) & (0.2541,0,0.0082) \\
V (4i) & (0.2523,0,0.0046) & (0.2535,0,0.0071) & (0.2540,0,0.2459) \\
V (4i) & (0.4965,0,0.2479) & (0.4935,0,0.2466) & (0.4920,0,0.2459) \\
 \hline
Sb (2a) & $\rightarrow$ & (0,0,0) & $\leftarrow$ \\
Sb (2b) & $\rightarrow$ & (0,1/2,0) & $\leftarrow$ \\
Sb (2c) & $\rightarrow$ & (0,0,1/2) & $\leftarrow$ \\
Sb (2d) & $\rightarrow$ & (0,1/2,1/2) & $\leftarrow$ \\
Sb (8j) & (0.1670,0.1305,0.3333) & (0.1672,0.1268,0.3331) & (0.1673,0.1221,0.3329) \\
Sb (8j) & (0.1663,0.3695,0.3333) & (0.1659,0.3732,0.3331) & (0.1656,0.3779,0.3329) \\
Sb (8j) & (0.3334,0.1318,0.1667) & (0.3334,0.1293,0.1667) & (0.3334,0.1251,0.1667) \\
Sb (8j) & (0.3332,0.3700,0.1664) & (0.3330,0.3735,0.1659) & (0.3328,0.3780,0.1655) \\
\end{tabular}
\end{ruledtabular}
\label{tab.latt12}
\end{table*}

\begin{figure*}[!tp]
\centering
\includegraphics[width=\linewidth]{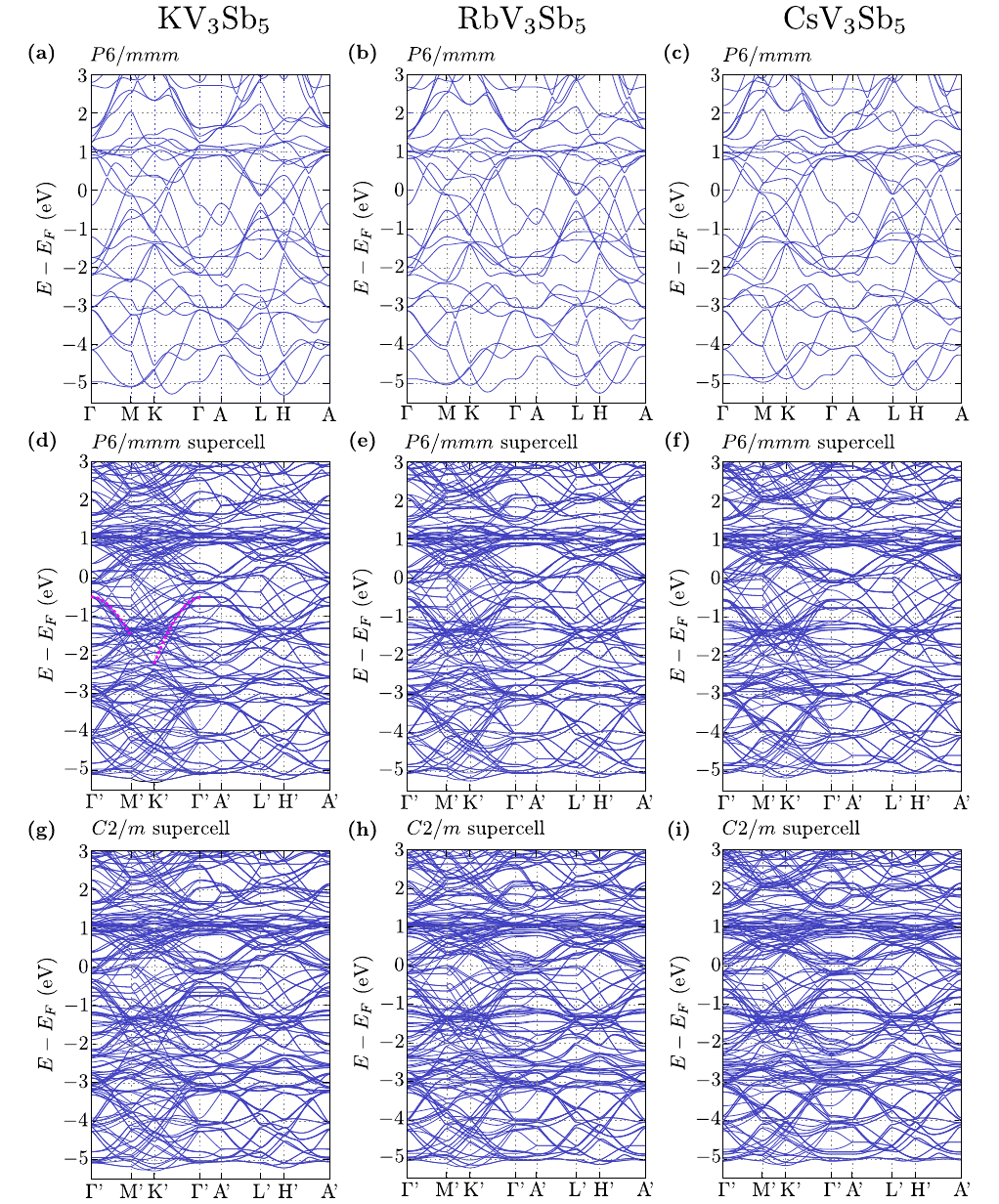}
\caption{
Electronic band structure of $A$V$_{3}$Sb$_{5}$ for different phases (columns from left to right correspond to $A =$ K, Rb, and Cs, respectively).
Rows from top to bottom correspond to phase $P6/mmm$ in the case of primitive cell and $2\times2\times2$ supercell, and $C2/m$ supercell (related to $2\times2\times2$ supercell with $P6/mmm$ symmetry), respectively.
\label{fig.el_band}
}
\end{figure*}

\begin{table*}[!t]
\caption{
The frequencies of the optical modes at the $\Gamma$ point (and their irreducible representations) with the indicated Raman (R) and infrared (Ir) activities for CsV$_{3}$As$_{5}$ in $P6/mmm$ and $C2/m$ phases.
}
\begin{ruledtabular}
\begin{tabular}{cl|clclcl}
\multicolumn{2}{c}{$P6/mmm$} & \multicolumn{6}{c}{$C2/m$} \\
 \hline
(THz) & IR & (THz) & IR & (THz) & IR & (THz) & IR \\
 \hline
1.505 & $E_\text{1u}$ (Ir) & 1.332 & $A_\text{u}$ (Ir) & 1.401 & $A_\text{g}$ (R) & 1.412 & $B_\text{u}$ (Ir) \\ 
1.763 & $A_\text{2u}$ (Ir) & 1.421 & $A_\text{g}$ (R) & 1.447 & $B_\text{u}$ (Ir) & 1.481 & $B_\text{g}$ (R) \\
2.047 & $B_\text{2u}$ & 1.502 & $B_\text{u}$ (Ir) & 1.509 & $B_\text{u}$ (Ir) & 1.515 & $B_\text{g}$ (R) \\
2.120 & $E_\text{1u}$ (Ir) & 1.519 & $B_\text{g}$ (R) & 1.547 & $B_\text{g}$ (R) & 1.577 & $B_\text{g}$ (R) \\
2.270 & $E_\text{1g}$ (R) & 1.624 & $B_\text{u}$ (Ir) & 1.657 & $A_\text{g}$ (R) & 1.740 & $B_\text{g}$ (R) \\
2.808 & $A_\text{2u}$ (Ir) & 1.759 & $A_\text{u}$ (Ir) & 1.790 & $A_\text{g}$ (R) & 1.791 & $A_\text{g}$ (R) \\
3.451 & $E_\text{2u}$ & 1.795 & $B_\text{g}$ (R) & 1.981 & $A_\text{u}$ (Ir) & 1.982 & $A_\text{u}$ (Ir) \\
3.967 & $E_\text{2g}$ (R) & 2.040 & $A_\text{u}$ (Ir) & 2.046 & $B_\text{u}$ (Ir) & 2.118 & $B_\text{u}$ (Ir) \\
4.194 & $A_\text{1g}$ (R) & 2.368 & $A_\text{u}$ (Ir) & 2.376 & $B_\text{g}$ (R) & 2.388 & $B_\text{g}$ (R) \\
4.478 & $B_\text{1g}$ & 2.484 & $A_\text{u}$ (Ir) & 2.485 & $A_\text{u}$ (Ir) & 2.729 & $A_\text{u}$ (Ir) \\ 
5.526 & $E_\text{1u}$ (Ir) & 2.799 & $B_\text{u}$ (Ir) & 2.829 & $A_\text{g}$ (R) & 2.862 & $B_\text{u}$ (Ir) \\
6.905 & $B_\text{2u}$ & 2.886 & $A_\text{g}$ (R) & 2.928 & $A_\text{u}$ (Ir) & 2.972 & $A_\text{u}$ (Ir) \\
7.621 & $E_\text{1u}$ (Ir) & 2.983 & $A_\text{u}$ (Ir) & 3.127 & $A_\text{g}$ (R) & 3.160 & $B_\text{u}$ (Ir) \\
7.656 & $A_\text{2u}$ (Ir) & 3.264 & $B_\text{u}$ (Ir) & 3.360 & $B_\text{g}$ (R) & 3.448 & $B_\text{u}$ (Ir) \\
8.892 & $B_\text{1u}$ & 3.449 & $B_\text{g}$ (R) & 3.454 & $B_\text{u}$ (Ir) & 3.462 & $B_\text{g}$ (R) \\
9.275 & $E_\text{2u}$ & 3.487 & $A_\text{u}$ (Ir) & 3.488 & $A_\text{u}$ (Ir) & 3.533 & $B_\text{g}$ (R) \\
& & 3.550 & $B_\text{g}$ (R) & 3.733 & $B_\text{u}$ (Ir) & 3.744 & $B_\text{g}$ (R) \\
& & 3.758 & $B_\text{u}$ (Ir) & 3.842 & $A_\text{g}$ (R) & 3.856 & $A_\text{g}$ (R) \\
& & 3.908 & $A_\text{g}$ (R) & 3.933 & $B_\text{u}$ (Ir) & 3.940 & $A_\text{u}$ (Ir) \\
& & 3.946 & $A_\text{g}$ (R) & 3.956 & $A_\text{g}$ (R) & 3.974 & $A_\text{u}$ (Ir) \\
& & 3.996 & $A_\text{u}$ (Ir) & 4.141 & $B_\text{u}$ (Ir) & 4.204 & $B_\text{u}$ (Ir) \\
& & 4.210 & $A_\text{g}$ (R) & 4.218 & $B_\text{u}$ (Ir) & 4.515 & $B_\text{g}$ (R) \\
& & 4.975 & $B_\text{u}$ (Ir) & 4.996 & $B_\text{u}$ (Ir) & 5.195 & $B_\text{u}$ (Ir) \\
& & 5.205 & $A_\text{g}$ (R) & 5.305 & $A_\text{g}$ (R) & 5.307 & $A_\text{g}$ (R) \\
& & 5.471 & $B_\text{u}$ (Ir) & 5.475 & $B_\text{u}$ (Ir) & 5.787 & $A_\text{g}$ (R) \\
& & 6.251 & $A_\text{g}$ (R) & 6.302 & $B_\text{u}$ (Ir) & 6.446 & $A_\text{g}$ (R) \\
& & 6.512 & $B_\text{u}$ (Ir) & 6.554 & $B_\text{u}$ (Ir) & 6.808 & $A_\text{g}$ (R) \\
& & 6.809 & $A_\text{g}$ (R) & 7.001 & $A_\text{g}$ (R) & 7.052 & $A_\text{g}$ (R) \\
& & 7.233 & $A_\text{g}$ (R) & 7.288 & $A_\text{g}$ (R) & 7.416 & $B_\text{u}$ (Ir) \\
& & 7.597 & $B_\text{u}$ (Ir) & 7.632 & $B_\text{u}$ (Ir) & 7.646 & $B_\text{u}$ (Ir) \\
& & 7.689 & $A_\text{u}$ (Ir) & 7.810 & $B_\text{g}$ (R) & 8.055 & $B_\text{g}$ (R) \\
& & 8.063 & $B_\text{u}$ (Ir) & 8.073 & $A_\text{u}$ (Ir) & 8.133 & $B_\text{g}$ (R) \\
& & 8.236 & $A_\text{u}$ (Ir) & 8.248 & $B_\text{u}$ (Ir) & 8.309 & $A_\text{u}$ (Ir) \\
& & 8.531 & $B_\text{g}$ (R) & 8.572 & $B_\text{g}$ (R) & 8.899 & $B_\text{u}$ (Ir) \\
& & 9.118 & $B_\text{g}$ (R) & 9.317 & $A_\text{u}$ (Ir) & 9.360 & $A_\text{u}$ (Ir) \\
\end{tabular}
\end{ruledtabular}
\label{tab.ir}
\end{table*}


\end{document}